\documentclass[11pt,a4paper,numbers]{article}
\pdfoutput=1
\usepackage{jcappub}

\usepackage{amssymb,amsmath,latexsym,graphics, graphicx,epsfig,multirow,comment,hyperref,appendix, verbatim} 
\usepackage{pifont}
\usepackage{lipsum}

\usepackage{slashed}
\usepackage{subfigure}
\usepackage{feynmp}
\usepackage{graphicx}
\usepackage{amsfonts}
\usepackage{color}
\usepackage{cancel}

\def\hbar{\hspace{0pt}\raisebox{1pt}{$-$} \hspace{-7pt} h}

\def\5{\overline 5}

\definecolor{JJ}{RGB}{0,144,255}

\newcommand{\be}{\begin{equation}}
\newcommand{\ee}{\end{equation}}
\newcommand{\bea}{\begin{eqnarray}}
\newcommand{\eea}{\end{eqnarray}}

\newcommand{\ba}{\begin{eqnarray}}
\newcommand{\ea}{\end{eqnarray}}

\begin{document}
\title{X-ray Lines from Dark Matter:\\ The Good, The Bad, and The Unlikely}
%

\author{Mads T. Frandsen,}
\author{Francesco Sannino,}
\author{Ian M. Shoemaker,}
\author{and Ole Svendsen}
\affiliation{CP$^{3}$-Origins \& Danish Institute for Advanced Study {DIAS}, University of Southern Denmark, Campusvej 55, DK-5230 Odense M, Denmark}
\arxivnumber{CP3-Origins-2014-007 DNRF90, DIAS-2014-7}

\abstract{

We consider three classes of dark matter (DM) models to account for the recently observed 3.5 keV line: metastable excited state DM, annihilating DM, and decaying DM.  We study two examples of metastable excited state DM. The first, millicharged composite DM, has both inelasticity and photon emission built in, but with a very constrained parameter space.  In the second example, up-scattering and decay come from separate sectors and is thus less constrained. 
The decay of the excited state can potentially be detectable at direct detection experiments. However we find that CMB constraints are at the border of excluding this as an interpretation of the DAMA signal.  The annihilating DM interpretation of the X-ray line is found to be in mild tension with CMB constraints. Lastly, a generalized version of decaying DM can account for the data with a lifetime exceeding the age of the Universe for masses $\lesssim 10^{6}$ GeV. 
}

\maketitle

\section{Introduction} 

We have yet to make a definitive non-gravitational detection of the dominant matter component in the Universe. Though we know quite well the cosmological abundance of this Dark Matter (DM), little else is known with certainty. Given that the first evidence for DM came from astronomical observations, it it is natural to hope we may continue to gain insights into its nature from such observations.

At present, a possible clue to the nature of DM may be coming from the recently reported 3.55 keV X-ray line. This line was first reported in the stacked analysis of 73 galaxy clusters~\cite{Bulbul:2014sua} with $>$ 3 $\sigma$ significance. A similar recent analysis finds evidence at the 4.4 $\sigma$ level for a 3.52 keV line from their analysis of the X-ray spectrum of the Andromeda galaxy (M31) and the Perseus cluster~\cite{Boyarsky:2014jta}.  The X-ray line has already generated considerable interest in the community as a possible signal of DM~\cite{Ishida:2014dlp,Finkbeiner:2014sja,Higaki:2014zua,Lee:2014xua,Jaeckel:2014qea,Abazajian:2014gza,Krall:2014dba}.
  
In both analyses (\cite{Bulbul:2014sua,Boyarsky:2014jta}), the unidentified line was interpreted as a possible signal of sterile neutrino dark matter (see e.g.~\cite{oai:arXiv.org:0906.2968}) decaying through a loop, $\nu_{s} \rightarrow \gamma + \nu$.  A complete Standard Model (SM) singlet, the sterile neutrino represents one of the simplest possibilities for DM, though DM could well be composed of richer structures. Here we explore three alternative models of DM that can account for the observed X-ray line: excited state DM, millicharged atomic DM, decaying DM, and we briefly comment on annihilating DM.

In section~\ref{basics} we estimate the basic requirements any model must satisfy to account for the observed X-ray flux. In section~\ref{milliDM} we examine a millicharged DM model with bound state formation.  In section~\ref{meta} we construct a simple model of pseudo-Dirac DM (i.e. fermionic DM with a large Dirac mass and small but nonzero Majorana mass) with a massive dark photon where the dominant coupling of DM to the dark photon is off-diagonal and connects two nearly degenerate states. For light mediator masses, this naturally leads to significant scattering, and therefore populates the excited state, inside galaxies and clusters. The excited state decays subsequently through a magnetic dipole to the ground state and an X-ray photon with energy equal to the mass-splitting.  In the final two sections, before offering our conclusions, we briefly consider annihilating DM and investigate is a generic model of dark matter decay.

\section{The Basic Requirements of the X-ray Signal}
\label{basics}

The observed 3 keV line is interpreted as the decay of a sterile neutrino, $\nu_{s} \rightarrow \nu + \gamma$ with mass 7.06 $\pm 0.5$ keV, with mixing angle $\sin^{2} 2 \theta = (2-20)\times 10^{-11}$,~\cite{Boyarsky:2014jta} i.e. an X-ray flux
\bea
\Phi_{X-ray} \propto n_{s} \Gamma_{s} &=& 1.39 \times 10^{-22} s^{-1} \sin^{2} 2 \theta \left(\frac{m_{s}}{{\rm keV}}\right)^{5} \rho_{DM}/m_{s} \\
&=& \left(1.5 \times 10^{-25}-2.7\times 10^{-24} \right)~ {\rm cm}^{-3} {\rm s}^{-1} ,
\eea
where we have used the DM density for Perseus, $\rho_{0}^{{\rm perseus}} = 0.03$ GeV ${\rm cm}^{-3}$~\cite{SanchezConde:2011ap}.

On the other hand, a class of inelastic scattering models with a wide range of DM masses can also accommodate the observed line. In such models, the rate of photon emission comes from the scattering rate of ground states into excited states, $\chi_{g} + \chi_{g} \rightarrow \chi_{e} + \chi_{e}$. Now in this case the X-ray flux has the scaling, $\Phi_{X-ray} \propto n_{\chi}^{2} \times (\sigma_{\chi_{g} + \chi_{g} \rightarrow \chi_{e} + \chi_{e}} v_{rel})\times {\rm BR(\chi_{e} \rightarrow \gamma + \chi_{g})}$, leading to
\be
 (\sigma_{\chi\chi} v_{rel}) \times BR \simeq (1.7 \times 10^{-22}-3.0\times 10^{-21})~{\rm cm}^{3}{\rm s}^{-1} (m_{\chi}/{\rm GeV})^{2}
\label{base}
\ee
In reasonable agreement with~\cite{Finkbeiner:2014sja}. Note that models of this type have been studied as a way to account for the anomalous 511 keV flux from the Galactic Center~\cite{Finkbeiner:2007kk}. 

The approximation we have made in the above is that scattering rate, rather than the decay rate is the bottleneck for the photon emission, i.e. $\Gamma_{\chi \chi} < \Gamma_{\chi_{e} \rightarrow \gamma + \chi_{g}}$. We shall see that for decay via a magnetic transition moment, $\bar{\chi}_{e} \sigma_{\mu \nu} \chi_{g} F^{\mu \nu}/\Lambda$ , this is satisfied for $\Lambda < 10^{14}~{\rm GeV}~\sqrt{{\rm GeV}/m_{\chi}}$, implying a lifetime $\lesssim 10^{21}s ~\left({\rm GeV}/m_{\chi}\right)$. 


Furthremore, let us comment on the complementary probes of this DM self-scattering. One probe of DM self-interactions is offered by the Bullet cluster~\cite{Randall:2007ph}, with the quoted limit $\sigma_{\chi \chi} /m_{\chi} \lesssim 1.8\times 10^{-24}~{\rm cm}^{2}/{\rm GeV}$. We see that the requirement that we simultaneously have sufficient scattering to account for the X-ray flux (Eq.~\ref{base}) while remaining consistent with the Bullet cluster requires simply:
\be 
1 ~{\rm GeV}\lesssim m_{\chi} \lesssim 1000~{\rm TeV},
\ee
where the lower bound simply comes the kinematic requirement that up-scattering to a 3.5 keV higher mass state can occur at cluster velocities. 

Lastly, it is important to note that the X-ray flux in either class of models is simply a product of the decaying particle's density $n_{e}$ and its decay rate $\Gamma_{e}$:
\be
\Phi_{X-ray} \propto n_{e}(t) \Gamma_{e} = \begin{cases} \frac{\rho_{\chi}}{m_{\chi}} \Gamma_{\chi}, & \mbox{decaying DM }\\(\sigma v_{rel}) \left(\frac{\rho_{\chi}}{m_{\chi}}\right)^{2}, & \mbox{metastable excited state DM}  \end{cases}
\ee
where the two classes denote the extreme cases where the decay rate or the scattering rate are dominant. More generally, when the rate of scattering and decay are comparable one can find X-ray fluxes proportional to some non-integer power of the DM density, $\Phi_{X-ray} \propto \rho_{\chi}^{n}$, where $1 \le n\le 2$. 

With these simple estimations in mind, let us proceed to a more detailed look at models which can account for the observed X-ray flux. 

\section{Millicharged Composite DM}
\label{milliDM}

We first assume that the keV line arises from a 2-body decay of an excited state of DM preserving the
symmetry responsible for the stability of the dark matter candidate or at least part of it. A
$Z_{2}$ symmetry is a minimal dark symmetry choice. If the DM ground state is a fermion $\chi$
the decay into a photon and another fermion must involve a chirality flip to preserve spin.
A candidate operator is a magnetic transition moment $\frac{1}{\Lambda}\bar{\chi}_e \sigma^{\mu\nu} \chi_g F_{\mu\nu}$. 
Instead if the DM ground state is a spin-0 particle the excited state must be spin-1 and vice versa if the DM ground state is spin-1 the excited state must be spin-0.

One example of nearly degenerate spin-0 and spin-1 states arises in the heavy-quark limit for hadrons. This limit was explored in the context of a model with DM comprised of dark mesons in \cite{Alves:2009nf}.
Another example is the hyperfine transition between spin-0 and spin-1 states of the hydrogen atom. This was explored in a model of dark hydrogen in \cite{Cline:2012is}. Here we will discuss the latter example as a candidate for the keV line.

We therefore consider a two component model, with a dark electron ${\bf e}$ and a dark proton ${\bf p}$ charged under a new, unbroken $U(1)$ gauge symmetry.  Interactions with the visible sector arise from the kinetic mixing between the dark and ordinary photon, $\varepsilon F_{\mu \nu} V^{\mu \nu}$. The atomic hydrogen-like bound states will be denoted ${\bf H}$.  Atomic DM models along these lines go back many decades~\cite{Goldberg:1986nk,Hodges:1993yb} and have started to receive renewed attention as of late~\cite{Kaplan:2009de,CyrRacine:2012fz,Cline:2012is,Cline:2013pca,Petraki:2014uza}.  
 
The millicharged variant of atomic DM was recently studied in~\cite{Cline:2012is}, where an unbroken $U(1)$ gauge symmetry gives rise to atomic bound states.  In the case that $m_{e} = m_{p}$, elastic scattering is suppressed with the lowest energy inelastic transition coming from hyperfine splitting. In general, the hyperfine splitting is 
\be
\Delta E_{hf} \equiv \delta = \frac{8}{3} \alpha_{X}^{4} \frac{m_{e}^{2} m_{p}^{2}}{(m_{e}+m_{p})^{3}},
\ee
and the binding energy of the atom is $B = \frac{1}{2} \alpha_{X}^{2} \mu_{\chi}$, where $\mu_{\chi} = m_{e}m_{p}/(m_{e} + m_{p})$ is the reduced mass. For DM 10 GeV range, this fixes the DM coupling to be $\alpha_{X} \sim 3 \times10^{-2}$.

\begin{figure*}[t!]
  \centering
                 \includegraphics[width=0.55\textwidth]{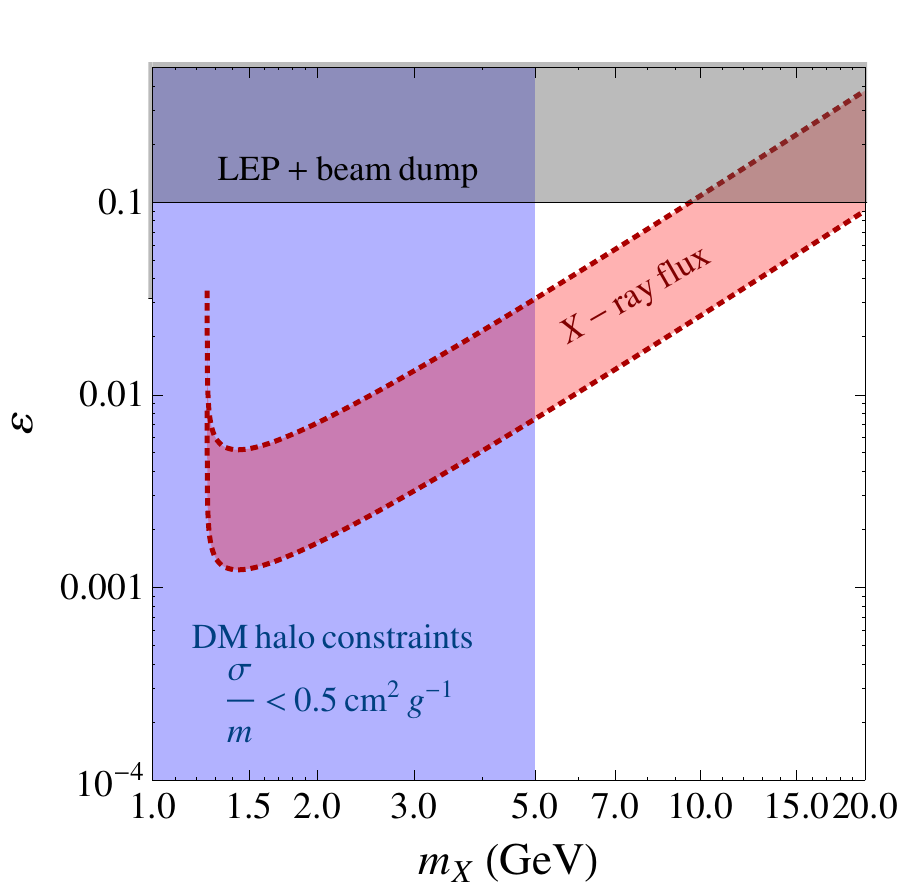}

   \caption{Here we plot the constraints from DM halo constraints (in blue) ruling out $\sigma/m_{\chi} < 0.5 ~{\rm cm}^{2}{\rm g}^{-1}$~\cite{Tulin:2013teo,Cline:2013pca}, along with the required scattering cross section needed to reproduce the correct X-ray flux (see Eq.~\ref{base}). The black region is excluded by the combination of LEP and beam dump searches~\cite{Akers:1995ui,Abreu:1990qu}, as reported in~\cite{Davidson:1991si}. Here we have fixed $m_{e} = m_{p} = m_{\chi}/2$. }
  \label{fig:milliDM}
\end{figure*}

The requisite self-scattering is achieved from dark hydrogen-dark hydrogen scattering~\cite{Cline:2013pca}
\be (\sigma_{{\bf H H}}v_{rel}) \simeq (\mu_{X} \alpha_{X})^{-2} \left[a_{0} + a_{1} \left(\frac{m_{\chi}v_{rel}^{2}}{ 4\mu_{\chi} \alpha_{X}}\right)+a_{2} \left(\frac{m_{\chi}v_{rel}^{2}}{ 4\mu_{\chi} \alpha_{X}}\right)^{2}\right]^{-1} \sqrt{v_{rel}^{2} - \frac{4\delta}{m_{X}}}
\label{eq:HH}
\ee
where the prefactors $a_{i}$ are determined from numerical fits (see their Table 1) to detailed calculation of the full transfer cross section.   The kinematic factor in Eq.~(\ref{eq:HH}) we include simply comes from the reduced phase space as the incoming kinetic energy approaches the mass-splitting, $\delta$.

The decay of the excited state proceeds dominantly via dark photon emission,  ${\bf H}^{*} \rightarrow {\bf H} + \gamma'$, since this is unsuppressed by the mixing $\varepsilon$, and to ordinary SM photons with a small branching ratio $BR({\bf H}^{*} \rightarrow {\bf H} + \gamma) \simeq \varepsilon^{2}\alpha/\alpha_{X}$. This two-body decay yields a photon of energy $E_{\gamma} = \Delta E_{hf}$. 

Given the small branching ratio, we require rather large scattering cross sections. We combine the requirement that $\Delta E_{hf} = 3.5$ keV and the scattering constraint Eq.~(\ref{base}) to illustrate the region of interest in Fig.~\ref{fig:milliDM}. Here we follow~\cite{Tulin:2013teo,Cline:2013pca} and impose $\sigma_{\bf H H}/m_{\chi} < 0.5$ cm$^{2}{\rm g}^{-1}$ at velocities 10-1000 km/s. A variety of constraints have been imposed on millicharged DM, ranging from White Dwarfs, Red Giants, Supernovae, BBN, and accelerator experiments. In the mass range we consider, only the accelerator searches are relevant, though the reader can find a detailed survey of the constraints in~\cite{Davidson:1991si,Davidson:2000hf,Abel:2008ai}.

The constraint on DM self-interactions from halo constraints requires DM to be $\gtrsim 5$ GeV in this example. (Note that in this benchmark example, the binding energy is $B\simeq 1$ MeV, and thus in typical collisions in clusters there is insufficient kinetic energy to ionize the atoms.) Moreover, given such a binding energy, we assume that the formation of electromagnetic bound states of DM with nuclei as considered in~\cite{Langacker:2011db} will not be efficient, though a detailed study of this possibility is beyond the scope of this work. 

In addition, at such light masses the scattering cross section with ordinary matter is only weakly constrained by direct detection experiments.  We find that relaxing the assumption of $m_{e} = m_{p}$ yields stronger constraints on the lower bound of the DM mass consistent with self-interaction limits. Given that this pushes the favored DM masses into a regime where direct detection limits are very significant, we do not explore this possibility further. 

Quite generally the dissipation of DM is strongly constrained. However in the absence of efficient molecular hydrogen ${\mathbf H}_{2}$ we expect the amount of DM dissipation to be small~\cite{Cline:2013pca}. 



Future tests of this model can come from low-threshold direction detection experiments, as well as higher energy photon lines from the additional atomic transitions.

\section{Metastable Magnetic DM}
\label{meta}

\begin{figure*}[t!]
  \centering
                 \includegraphics[width=0.55\textwidth]{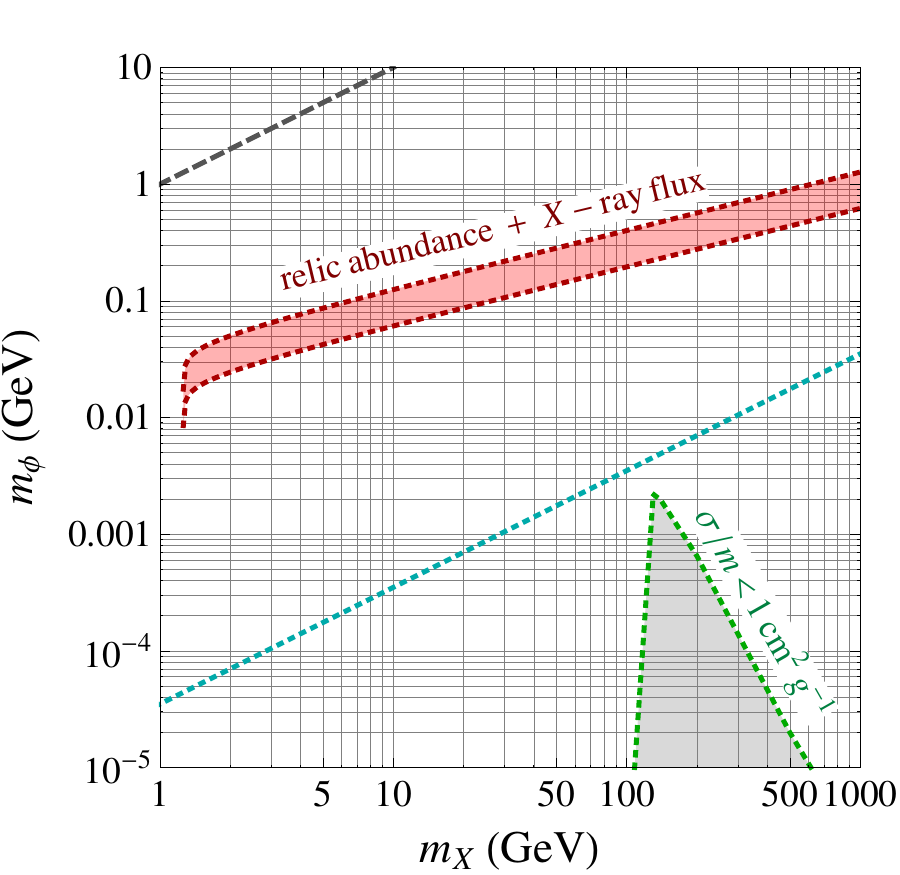}

   \caption{Here we plot the required $(\alpha_{X}, m_{\chi})$ for achieving the correct relic abundance and the needed scattering cross section for the observe X-ray flux. Our perturbative calculation breaks down when, $\alpha_{X}m_{X}/m_{\phi} \simeq 1$, as indicated by the blue dashed curve. For reference, the dashed gray line indicates where DM annihilation to $\phi \phi$ becomes kinematically inaccessible.  The gray region below the green curve indicates parameter points excluded by the requirement that $\sigma_{\chi\chi}/m_{\chi} < 1~{\rm cm}^{2}{\rm g}^{-1}$ on galactic scales~\cite{Vogelsberger:2012ku,Rocha:2012jg,Peter:2012jh}.}
  \label{fig:meta}
\end{figure*}
We take DM to be fermionic with Dirac mass $M$ and Majorana mass $ \delta \ll M$. In terms of the interaction eigenstate Dirac fermion $\chi$ we can write the mass terms as,  $M\bar{\chi}\chi + \delta (\chi_2 \chi_2 + \chi_1 \chi_1+  h.c.)$ where $\chi_{1,2}$ are Weyl fermions, $\chi= (\chi_1, \chi_2^c)$.
In addition, we assume that the DM $\chi$ (interaction eigenstate) is coupled to a new gauge field $\phi_{\mu}$ of a spontaneously broken gauged $U(1)_{X}$. This new gauge field will in general mix kinetically with the ordinary photon,  $\varepsilon F_{\mu \nu} V^{\mu \nu}$, and thereby provide a connection to SM fields.  Now one can diagonalize the mass matrix and write the Lagrangian in terms of mass eigenstates $\chi_{g,e}$ with $\chi_g$ the lightest state 

\be \mathcal{L}_{\rm int} = \overline{\chi} i \gamma^{\mu} \left( \partial_{\mu} + i g_{X} \phi_{\mu} \chi\right) \longrightarrow
\overline{\chi}_{g} i \gamma^{\mu}  \partial_{\mu}\chi_g  + \overline{\chi}_{\,e} i \gamma^{\mu}  \partial_{\mu}\chi_e + (i g_{X} \phi_{\mu} \overline{\chi}_{\,g} \gamma^{\mu}\chi_e + {\rm h.c.}) + \mathcal{O}\left(\frac{g_X \delta }{M}\right), \nonumber \\
\ee
where the mass eigenstates are $M_{g,e} = M \mp \delta$ and $\delta$ is assumed to be positive. From the above expression, it is thus clear that the interaction is dominantly off-diagonal.~\footnote{The next term in the $m/M$ expansion is diagonal but spin-dependent.}

At small $\phi$ mass the DM annihilation cross section is dominated by $\chi_{i} \chi_{i} \rightarrow \phi \phi$ where 
\be \langle \sigma_{\chi_{i} \chi_{i} \rightarrow \phi \phi} v _{rel}\rangle = \frac{\pi \alpha_{X}^{2}}{m_{\chi_{i}}^{2}} \sqrt{ 1- \left(\frac{m_{\phi}^{2}}{m_{\chi_{i}}}\right)^{2}}
\ee
Thus achieving the right relic abundance requires $\alpha_{X} \gtrsim 3 \times 10^{-5} (m_{\chi}/ {\rm GeV})$, where the inequality is saturated for a symmetric species.

The self-scattering rate of ``double up-scattering'' is estimated simply as
\be (\sigma_{\chi_{g} \chi_g \rightarrow \chi_{e} \chi_{e}} v_{rel}) = \frac{4 \pi \alpha_{X}^{2} m_{X}^{2}}{m_{\phi}^{2}(m_{\phi}^{2} + m_{\chi}^{2} v_{rel}^{2})} \sqrt{v_{rel}^{2} - 4 \delta /m_{\chi}}
\ee
%

\begin{figure*}[t!]
  \centering
                 \includegraphics[width=0.45\textwidth]{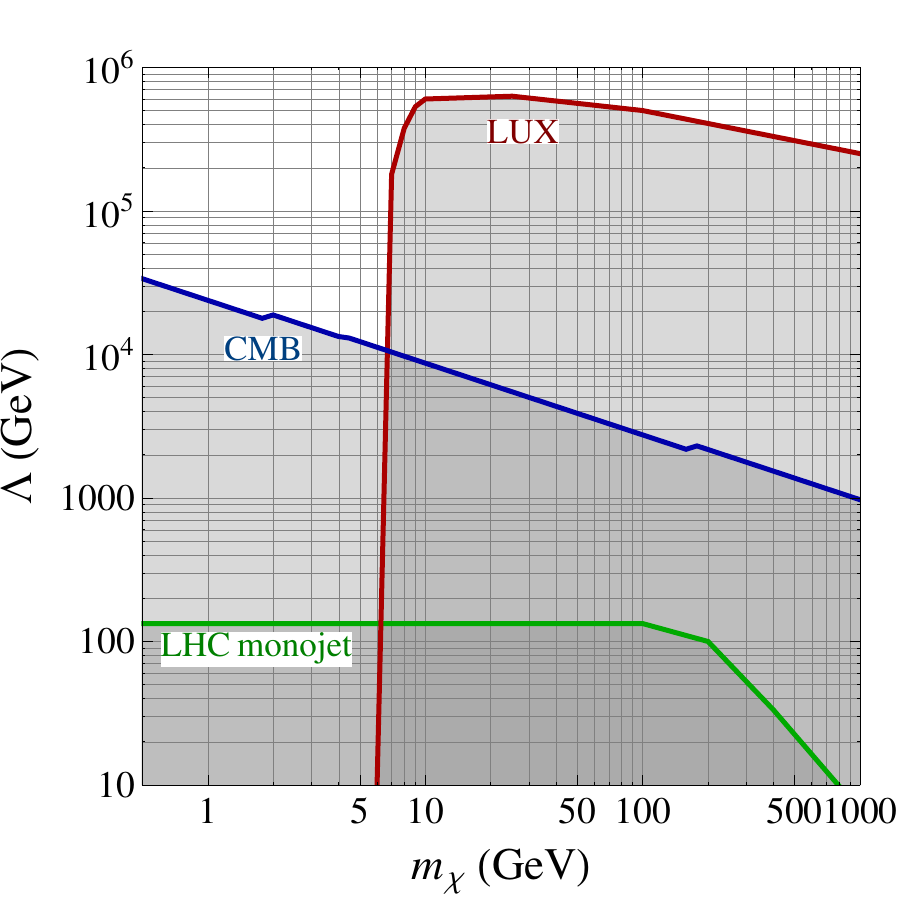}

   \caption{Constraints on the magnetic dipole operator. The LUX constraints are derived with a mass splitting $\delta =3.5$ keV. For the up-scattering interpretation of the X-ray signal, the decay rate must be larger than scattering rate, implying $\Lambda \lesssim 10^{15}$ GeV, being easily satisfied by the depicted constraints.}
  \label{fig:dipole}
\end{figure*}

Lastly once the excited states are sufficiently populated by the self-scattering, they de-excite to the ground state via the transition dipole moment, $\bar{\chi}_{e} \sigma_{\mu \nu} \chi_{g} F^{\mu \nu}/\Lambda$ by emitting an x-ray photon. This proceeds with the rate
\be \Gamma_{\chi_{e} \rightarrow \gamma + \chi_{g}} = \frac{4\delta^{3}}{\pi \Lambda^{2}}.
\label{eq:decay}
\ee

The final ingredient is to estimate the branching of the excited state into the two-body final state, ${\rm BR}(\chi_{e} \rightarrow \chi_{g} + \gamma)$. Branching ratios $\mathcal{O}(1)$ are quite natural for such small mass-splittings. The only other decay modes available arise from the kinetic mixing with the $Z$ and photons, allowing $ 3 \gamma$ (through charged fermions loops) and $\bar{\nu}\nu$ (from mixing with the $Z$) decay modes.  Even for relatively large values of $\varepsilon$ these decays are highly suppressed~\cite{Batell:2009vb,Pospelov:2008jk} compared to the $ \gamma+ \chi_{g} $ mode. Let us lastly note that the cosmology of light, kinetically mixed vector mediators has been argued to require $\varepsilon \gtrsim 10^{-10}$ in order for $\phi_{\mu}$ to decay before BBN~\cite{Lin:2011gj}. 


Of course, the DM self-scattering cannot be arbitrarily large~\cite{Feng:2009hw,Loeb:2010gj,Tulin:2013teo,Kahlhoefer:2013dca}. The constraints on DM self-interactions are typically quoted as limits on the transfer cross section, $\sigma_{T} = \int d \Omega (1- \cos \theta) d \sigma / d\Omega$. In particular we impose $\sigma_{\chi\chi}/m_{\chi} < 1~{\rm cm}^{2}{\rm g}^{-1}$ on galactic scales~\cite{Vogelsberger:2012ku,Rocha:2012jg,Peter:2012jh}. For inelastic scattering, the constraint is most relevant at high DM masses where non-perturbative effects become relevant. In addition, nearly as soon as the process becomes kinematically allowed the dependence on $\delta$ disappears. We therefore use the analytic results have been obtained in the classical limit ($m_{X}v/m_{\phi} >1$), where the scattering proceeds as \cite{Feng:2009hw,PhysRevLett.90.225002},
\begin{equation}
\sigma_T \approx \left\{ \begin{array}{cc} \frac{4\pi}{m^2_\phi} \beta^2 \ln(1+\beta^{-1}),
& \beta<0.1, \\
\frac{8\pi}{m^2_\phi}
\beta^2/(1+1.5\beta^{1.65}),
& 0.1 \le \beta \le 1000, \\
\frac{\pi}{m^2_\phi}
\left( \ln\beta + 1 - \frac{1}{2} \ln^{-1} \beta \right)^2,
& \beta > 1000,
\end{array} \right.
\end{equation}
where $\beta \equiv 2 \alpha_X m_\phi / (m_\chi v_\mathrm{rel}^2)$.

The superposition of both the halo constraints on self-interactions and the need X-ray flux is shown in Fig.~\ref{fig:meta}. There we see that in contrast to the millicharged atomic DM case, here the model easily evades the constraints on self-interactions. Note that the parameter space that the X-ray signal favours does not produce interesting effects in dwarf galaxies along the lines of~\cite{Loeb:2010gj,Tulin:2013teo}. We find that to simultaneously have an impact on dwarf galaxies while accounting for the X-ray flux would require $\gtrsim 10$ TeV DM and a small branching ratio into photons, $BR(\chi_{e} \rightarrow \gamma + \chi_{g})\simeq 10^{-9}$. 



The magnetic transition dipole moment could arise from a Yukawa coupling of the form $\mathcal{L}=  S \bar{E}\chi y+ h.c$, along with a Majorana mass for $\chi$ as above, where $E$ is a new heavy electron and $S$ is a new charged scalar and $y$ is a Yukawa coupling, such as in the $S \bar{E}\chi y$ model~\cite{Dissauer:2012xa,Frandsen:2013bfa}. In itself the $S \bar{E}\chi y$ model only  allows for the correct thermal relic density of dark matter for relatively heavy DM mass~\cite{Frandsen:2013bfa}. However augmenting the model by the above vector boson allows both the correct relic density and the requisite up-scattering cross section.

\subsection{Phenomenology of Magnetic Inelastic DM}
Magnetic inelastic dark matter can be probed in a number of ways: (1) through (endothermic) up-scattering off the nucleus into the excited state~\cite{Chang:2010en,Lin:2010sb,Patra:2011aa,Weiner:2012gm}, (2) from (exothermic~\cite{Batell:2009vb,Graham:2010ca,McCullough:2013jma,Frandsen:2014ima}) down-scattering of the excited state, and (3) through photon emission from decay of the excited state in the detector~\cite{Feldstein:2010su,Pospelov:2013nea}. 

This final possibility was proposed in the ``luminous dark matter'' scenario~\cite{Feldstein:2010su} to accommodate the DAMA results, while remaining consistent with the null results of the liquid xenon experiments. However the combination of constraints from the CMB~\cite{Galli:2011rz}, LHC monojets~\cite{Barger:2012pf}, and LUX appear to rule out the possibility of terrestrially excited DM decaying inside a direct detection experiment, unless $m_{X} \lesssim 10$ GeV, see Fig.~\ref{fig:dipole}. For large DM masses our LUX limits agree well within those already obtained in the elastic limit~\cite{Frandsen:2013bfa,Vecchi:2013iza,DelNobile:2014eta}.


After up-scattering, the excited state travels on average a distance
\be \ell_{D} = \frac{v}{\Gamma} \simeq 270~{\rm km}~\left(\frac{\Lambda}{10~{\rm TeV}}\right)^{2}~\left(\frac{3.5~{\rm keV}}{\delta}\right)^{3}.
\ee
Thus in view of the dipole constraints depicted in Fig.~\ref{fig:dipole}, DM $\lesssim 8$ GeV can scatter in the Earth and be sufficiently long-lived to lead to novel direct detection prospects along the lines considered in the ``luminous DM'' proposal~\cite{Feldstein:2010su,Pospelov:2013nea}.

To account for DAMA however, the analysis of~\cite{Feldstein:2010su} found that DM masses in the range $m_{X} \simeq 0.8-0.9$ GeV and dipole $\Lambda \simeq 7-23$ TeV range were necessary. This is on the border of our estimate of the CMB constraint, which requires $\Lambda \gtrsim 25$ TeV at 0.9 GeV.

Additional (weaker) constraints not depicted include the Fermi-LAT annihilation limits from nearby dwarf galaxies~\cite{Ackermann:2011wa}, gamma-ray line searches, and large-scale structure limits~\cite{Dvorkin:2013cea}.



\section{Annihilating DM}

In this case, the process $\chi \bar{\chi} \rightarrow \gamma \gamma$ can account for the observed X-ray excess when $m_{\chi} \simeq 3.5$ keV. For DM this light however, it can contribute to the effective number of neutrinos $N_{eff}$. It has been recently argued that DM annihilating into photons is safe for sub-MeV masses only in the case of DM as a real scalar~\cite{Boehm:2013jpa}, simply because it is only this case that is within the uncertainties of $N_{eff}$ limits.

From Eq.~(\ref{base}) the requisite annihilation cross section is simply estimated as:
\be \langle \sigma_{\chi \bar{\chi} \rightarrow \gamma \gamma} v_{rel} \rangle \simeq (2\times10^{-33}-4\times10^{-32})~{\rm cm}^{3}~{\rm s}^{-1} 
\label{ann}
\ee

For $s$-wave annihilation, this cross section is at odds with strong CMB constraints on injecting energy into the electromagnetic bath at late times. In particular requiring~\cite{Galli:2011rz,Lin:2011gj}
\be \frac{\langle \sigma_{\chi \bar{\chi} \rightarrow \gamma \gamma} v_{rel} \rangle_{CMB} }{m_{\chi}}< \frac{2.42 \times 10^{-27}~{\rm cm}^{3}~{\rm s}^{-1}}{{\rm GeV}}
\ee
or in other words, for $m_{\chi} = 3.5$ keV,  $\langle \sigma_{\chi \bar{\chi} \rightarrow \gamma \gamma} v_{rel} \rangle_{CMB}< 8.5 \times 10^{-33}~{\rm cm}^{3}~{\rm s}^{-1}$. Thus the constraint from the CMB constrains $s$-wave annihilating scenarios to lie at the lower end of the possible annihilation cross section, Eq(\ref{ann}).

{\it Prima facie}~ DM annihilating to photons via a $p$-wave cross section is also possible. The constraints from $N_{eff}$ require such light DM to be a real scalar~\cite{Boehm:2013jpa}. However the simplest tree-level completions of this type contain vector mediators, which are forbidden to decay into a pair of vectors by the Landau-Yang theorem~\cite{Landau:1948kw,Yang:1950rg}. We therefore conclude that it is challenging for simple, tree-level models of annihilating DM to account for the observed X-ray flux. We note that~\cite{Dudas:2014ixa} reaches similar conclusions.


\begin{figure*}[t!]
  \centering
                 \includegraphics[width=0.45\textwidth]{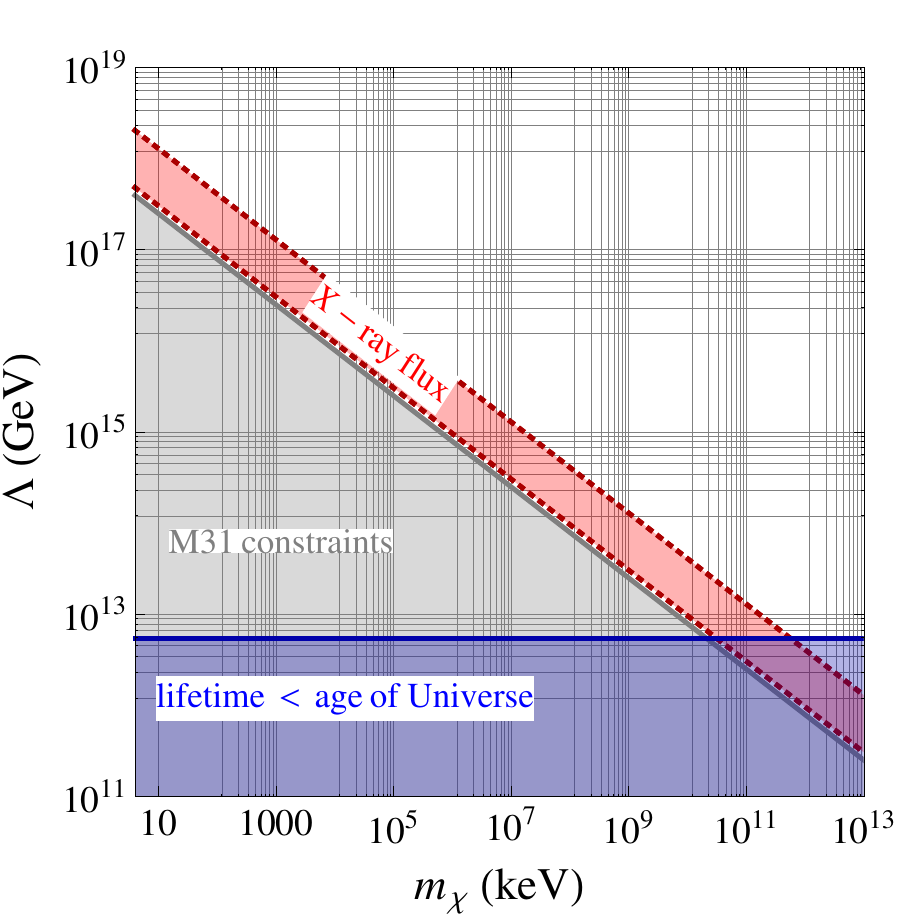}

   \caption{Here we consider decaying DM via $\frac{1}{\Lambda}\bar{\chi}_{e} \sigma_{\mu \nu} \chi_{g} F^{\mu \nu}$ and illustrate the required magnetic dipole scale $\Lambda$ and DM mass $m_{\chi}$ to account for the X-ray line while remaining consistent with the null results from M31~\cite{Watson:2006qb,Abazajian:2006jc}. }
  \label{fig:decay}
\end{figure*}

%
\section{Decaying DM}
Another simple model that can account for the line, is slowly decaying dark matter. The sterile neutrino is one example of this. However more generally, the line could simply arise from the decay of DM finally reaching the ground state today. One possibility for this scenario is the metastable magnetic scenario described in Sec.~\ref{meta} with the modification that the decay rate is much smaller than the up-scattering rate. Other possibilities for proceeding a large population of excited DM states also exist~\cite{Batell:2009vb,Graham:2010ca}.  With this in mind we will proceed to explore the possibility that DM is slowly decaying via, $\bar{\chi}_{e} \sigma_{\mu \nu} \chi_{g} F^{\mu \nu}/\Lambda$. 

Now we simply need $n_{X} \Gamma_{\chi_{e} \rightarrow \chi_{g} + \gamma}$ to be comparable to the sterile neutrino case.  The resulting requirement is shown in Fig.~\ref{fig:decay}, which was produced using Eq.~(\ref{eq:decay}). In addition we show the limits from M31 X-ray line searches~\cite{Watson:2006qb,Abazajian:2006jc}, which are not constraining. For DM not to decay on a timescale less than the age of the Universe while accounting for the X-ray signal as in Eq.(\ref{base}), the DM mass must be $\lesssim 10^6$ GeV. Intriguingly, a dipole scale $\Lambda$ consistent with the X-ray line can coincide with the GUT or see-saw scale of neutrino masses.  

Moreover note that the DAMA signal cannot be accommodated by decaying DM.  Let us take the sterile neutrino to begin with. Following~\cite{Ando:2010ye}, the event rate can be estimated as $\sim10^{-10}$ per cubic meter per year with the parameters for the X-ray signal~\cite{Boyarsky:2014jta}.  Looking at Fig.~\ref{fig:decay} we see that one cannot gain sufficiently in the event rate to get any appreciable signal in a $\sim$0.1 ${\rm m}^{3}$ detector such as DAMA.

\section{Conclusions}
We have studied up-scattering, annihilating and decaying DM models that can accommodate the recently reported 3.5 keV X-ray line. The up-scattering example of miilicharged atomic model is consistent with the limits of self-scattering in the Bullet cluster and direct detection. We consider a similar model of excited state fermionic DM with a vector mediator and a magnetic dipole. At present this model is the least constrained, and most straightforward to accommodate with the relic density requirements. Further tests of both scattering models can come from improved statistics on the radial distribution of the signal, making clear predictions that deviate from the decaying DM scenario. More generally a metastable model predicts, $\Phi_{X-ray} \propto \rho(r)^{n}$ where $1\le n \le 2$, where $n$ depends on the relative scattering and decay rates.  As emphasised previsouly~\cite{Finkbeiner:2014sja} this class of models also predicts higher X-ray fluxes from clusters than from galaxies or dwarfs, as well as photon signals in direct detection experiments~\cite{Feldstein:2010su,Pospelov:2013nea}.  Lastly, a novel probe of the lifetime of the excited state could come from a 3.5 keV excess in clusters which have recently undergone collisions. We hope to return to this topic in future work.

\section*{Note Added}
After our work appeared the paper~\cite{Dudas:2014ixa} also studied models of annihilating DM to account for the X-ray flux. They reach similar conclusions to ours, finding that the CMB constraints are in mild tension and that a viable model requires scalar DM with a higher-dimensional operator coupling to photons.

\section*{Acknowledgements}

We would like to thank Jim Cline, Felix Kahlhoefer, Chris Kouvaris, Kallia Petraki, Kai Schmidt-Hoberg, and Jussi Virkajarvi for helpful discussions.
The CP$^3$-Origins centre is partially funded by the Danish National Research Foundation, grant number DNRF90.  MTF acknowledges a Sapere Aude Grant no. 11-120829
from the Danish Council for Independent Research.

\bibliographystyle{JHEP}
\bibliography{nu.bib}

\end{document}